# Development and Experimentation of a Software Tool for Identifying High Risk Spreadsheets for Auditing

Mahmood H. Shubbak, Simon Thorne
Cardiff Metropolitan University
sthorne@cardiffmet.ac.uk

**ABSTRACT**

*Heavy use of spreadsheets by organisations bears many potential risks such as errors, ambiguity, data loss, duplication, and fraud. In this paper these risks are briefly outlined along with their available mitigation methods such as: documentation, centralisation, auditing, and user training. However, because of the large quantities of spreadsheets used in organisations, applying these methods on all spreadsheets is impossible. This fact is considered as a deficiency in these methods, a gap which is addressed in this paper.*

*In this paper a new software tool for managing spreadsheets and identifying the risk levels they include is proposed, developed and tested. As an add-in for Microsoft Excel application, 'Risk Calculator' can automatically collect and record spreadsheets properties in an inventory database and assign risk scores based on its importance, use, and complexity. Consequently, auditing processes can be targeted to high risk spreadsheets. Such a method saves time, effort, and money.*

## 1. BACKGROUND

Financial, commercial, industrial, service sector, government, construction, and educational institutions report high levels of spreadsheet use [SERP, 2006]. This high use is no surprise since spreadsheets tend 'fill the gap' between IT system capabilities and current business requirements [Chambers & Hamill, 2008]. However, this large scale uncontrolled use of spreadsheets exposes these organisations to significant risks.

**1.1. Spreadsheet Risks**

Spreadsheet risks can be categorised into: Data Loss; Errors; Ambiguity; Conflicts and Redundancy and Fraud.

### 1.1.1    Data Loss

When spreadsheets are used for critical decision-making, such as calculating the company's budget, supporting strategic reports, or negotiating deals with clients, some important questions should be asked; what would be the case if some of these critical sheets were lost? What are the consequences on the organisation's deals with clients and their confidence? What are the implications on the validity and accuracy of the decisions based on incomplete information?

### 1.1.2    Errors

According to Panko [1998]:

> 'There has long been ample evidence that errors in spreadsheets are pandemic. Spreadsheets, even after careful development, contain errors in 1% or more of all formula cells. In large spreadsheets with thousands of formulas, there will be dozens of undetected errors'

Panko [1996] demonstrated that with even the simple task of calculating the costs of building a wall, 38% of participants from MIS background have made errors. Field studies of errors show that spreadsheets in industry carry a significant number of errors too, some studies reporting up to 90% of the models audited carrying at least one material error

Errors can be introduced to spreadsheets at any time and in various different forms, they can occur during the manual entry of data, during internal or external linking, while writing formulas, or within copy-paste operations.

Spreadsheets are usually designed and used by employees with no formal training sessions [Bakeret al, 2006], Spreadsheets that support critical decision making are susceptible to the 'errors risk'. Errors here mean wrong decisions at the operational level, which can affect the efficiency of organisation's day to day activities. On the other hand, wrong estimations of assets, costs, or taxes can directly affect the strategic level.

### 1.1.3 Ambiguity

Using Excel without a systematic procedure of development and documentation can lead by time to ambiguity, especially when these files are used by many users. Misunderstanding of the sheet content, graphs, variable names and formulas can lead to handling them incorrectly, either at the data entry level or in generating reports. If these reports are then used by management in decision making, it will entail serious consequences.

### 1.1.4 Conflicts and Redundancy

Owing to their flexibility, it is likely that users prefer designing new Excel sheets than searching for similar archived ones. This attitude ends up, in many cases, with a large number of redundant sheets. Sometimes conflicts between those sheets exist; making it very difficult for users to find the most accurate information and integrate all these sheets together [Hermans, 2012].

This risk can delay the processing time and result in inaccurate reports. Moreover, these redundant sheets usually occupy high portions of the organisations' servers, reducing their efficiency and information security as well.

### 1.1.5 Fraud

The uncontrolled large use of Excel sheets can negatively affect companies' security, "They can be opening up opportunities for fraud" [Denizon, 2012], critical information can fall into the wrong hands, and large losses can incurred consequently.

In one of the 'horror stories', Allied Irish Bank (AIB) reported in March 2002 a large loss of approximately $700 Million in a fraud [Thorne, 2013]. The fraud committed was hidden by spreadsheets that were used to assess risk.

**1.2. Risk Mitigation Methods**

Existing approaches to mitigating spreadsheet risk begin usually with defining the main factors behind those risks. Accordingly, two different sets of solutions emerge in the literature either based on the view that spreadsheet risk arises from technical issues or that spreadsheet risk arises from management practices. Hence the solutions broadly fall into these two categories either reduce risk through technical or managerial solutions.

**1.2.1 Technical solutions**; relate to the software itself in terms of its interface and characteristics, under which the following solutions can be found:

   a. Altering spreadsheet environments [Paine, Ensuring Spreadsheet Integrity with Model Master, 2001]

   b. Methodology and Process [Paine et al., 2006][Grossman & Özlük, 2004]

   c. Centralisation; either in a common server [Informatica, 2012], or using Cloud Computing [Mell & Grance, 2011] however, this might encounter many risks, that sometimes exceed the spreadsheet risks [Armbrust, et al., 2009].

The main issue with technical solutions is that by significantly altering the spreadsheet environment, the solution is far less likely to be applicable to the artefacts in most organisations. Hence whilst inventing a new environment might be applicable in niche situations, it is unlikely to be universal before taking into consideration the cost and other barriers to adoption.

**1.2.2  Managerial approaches**; related to:

   a. Internal policy and strategy [Chambers & Hamill, 2008]

   b. Training sessions [Stott, 2013].

   c. Best Practice [Colver, 2004]

   d. Auditing [Panko, 1998].

Although the underlying process described by [Chambers and Hamill, 2008] may provide a framework for management solutions to spreadsheet risk in organisations, the detail of such plans is highly sensitive to the organisation. As [Colver, 2004] points out, 'best practice' in one situation can exacerbate problems in another. Hence the same issue arises with management approaches, a certain set of best practices might apply well in one organisation but are rarely universal.

**1.3 The best defence**

By far the most effective method of limiting spreadsheet errors and subsequently risk is through auditing and code inspection activities. Research shows this to be the most effective approach to catching and correcting errors, with group auditing catching two thirds of mistakes [Panko, 1998].

However, the spreadsheet problem is so large that it is not feasible to audit every single model in an organisation. Automated spreadsheet auditing tools can streamline and assist in such activities but this software still needs a skilled auditor to make decisions and to infer if mistakes have been made.  OPERIS Analysis Kit [OPERIS, 2012], Spreadsheet Detective [Berglas, 2001], Spreadsheet Studio [McDaid, 2011], Rainbow Analyst [Shallcross, 2001], and SpACE [AuditWare, 2013], are examples of such tools. Whilst these tools are designed to audit one spreadsheet at a time in depth, they do not generally offer a means of assessing the riskiness of a spreadsheet (or group of spreadsheets) to assist in deciding which spreadsheet models to audit.

What follows is an overview of the tool presented in this paper which attempts to answer the problem of sorting spreadsheets into risk categories so that auditing processes can be directed more efficiently. The software uses a number of separate and customisable metrics based on importance, complexity and usage characteristics.

## 2.0 RISK CALCULATOR

The risk calculator is an add in tool for Microsoft Excel that firstly scans and builds an inventory of spreadsheets on an organisations server or network, calculates absolute and relative risk scores. The process is as follows:

1. Firstly, the software scans the main characteristics of the available spreadsheets within the organisation. This information is obtained from Excel and through user interaction.

2. Depending on the scan results, a filtering and sorting process will follow; in which some spreadsheets are suggested to be migrated into another software applications that are more suitable, while other spreadsheets will simply be ignored.

3. Finally a risk quantification procedure is calculated. Each spreadsheet is given two different risk scores; the first is an absolute score, while the second is a relative score comparing to other spreadsheets available in the organisation.

After that, the high-risk spreadsheets can be identified and appropriate procedures for risk reduction can be instigated. Moreover, a curriculum for training sessions as well as special purpose templates can be designed in accordance with their nature and content based on the findings of the process.

### 2.1 Risk Indicators

The following section discusses the various risk indicators that are used to build a risk scorecard for the risk calculator application

### 2.1.1 Spreadsheet Nature

Inside a firm's servers and computers, one can find large numbers of Excel files that have already reached the end of their life cycle, i.e. they already achieved the purpose they have originally been created for. These old files are thus not being used anymore in the company's daily work, neither for maintaining data, nor for processing it. Identifying these files from the initial stages of the risk management policy is important in order to exclude them from any further consideration. This saves time and processing power in the following steps.

On the other hand, the risk level of a spreadsheet is directly related to the nature of its content. Spreadsheets can either be used for maintaining and observing data, or for processing and doing calculations on it. As calculating spreadsheets include data input, calculation, and output processes, it is obviously more susceptible to human errors either in designing the equations, or in the data entry stages [Panko, 1998].

Those two points are the main factors of consideration under the spreadsheet nature, however to complete the image, an additional piece of information regarding the using category of the sheet is also required. What we mean by category is the group to which the main activity of the sheet is related, e.g. whether it is mainly related to employees, customers, assets, projects, or products and services. Knowing the spreadsheet category can facilitate filtering and other file management processes.

In order to understand and estimate the risk within a spreadsheet, we considered the following three main factors;

### 2.1.2 Importance

Importance in Risk Calculator is measured via three different dimensions; financial, strategic, and security related.

Financial importance is based on the approximate amount of money included in the spreadsheet. Since each Excel sheet can contain more than one currency formatted cell, which are sometimes totals and subtotals of other cells, the approach chosen for money approximation here is to find the maximum value within all currency formatted cells. The maximum here can give a more effective indication to the financial importance of the sheet, avoiding the exaggerated results of the summation operations.

The second consideration is given to the strategic importance of the Excel sheet, whether it is used for generating reports that support decision making processes or not. Such an indicator should have a significant share in identifying critical spreadsheets in the firm's business.

The third and last importance consideration is given to the safety and security relatedness of the spreadsheet. Excel sheets are sometimes used for holding some sensitive data, these sheets however might not have high monetary or direct strategic decision making relations, but containing such critical information, they definitely need to be handled more carefully. For this reason, the risk calculator add-in takes this indicator into consideration too.

### 2.1.3 Complexity

While the humans' learning process is accumulative by its nature, human's errors seems to be accumulative too. The more you do, the more serious errors you can make. Consequently, it can be predicted that large spreadsheets with more cells, formulas, variables and charts, encounter more end-user programming risk. These factors are used in the Risk Calculator add-in along with some other deeper measurements as spreadsheet complexity indicators.

Very little research has been conducted on whether the use of VBA in spreadsheet models is as error prone as actual spreadsheet modelling, i.e. do we find the same level of mistakes in VBA as in the spreadsheet model itself. However, one can assume that the inclusion of VBA is likely to make the model more complex and more difficult to audit. Similarly, nested if statements are usually a source of errors and ambiguity risks [Hermans, 2014].

The Solver add-in in Microsoft Excel is normally used to solve optimization problems. Such problems are mainly related to operations research and decision making. Besides its complexity indication it also shows the strategic importance of the sheet. Therefore, it can be considered as a good indicator of spreadsheet risk.

### 2.1.4 Spreadsheet Use

As explained in section 1.1, when the sheet is used and modified by many users, it becomes more susceptible to errors and ambiguity risks. The more users using a spreadsheet, the larger risk it could have, because of the possible dynamic changes on the spreadsheet they can do.

Moreover, being used frequently, even for solely data reading purposes, can indicate that the sheet is important somehow for the firm's daily operations.

Finally, spreadsheets that are transmitted externally are usually with a special importance. Such transmitting indicates that the sheet is accessed, used, and might be modified by different persons from different organizations, which increases the potential risk associated with it [Chambers & Hamill, 2008].

**2.1.5 Summary of Risk Indicators**

The following risk indicators are used in the calculation of risk of a spreadsheet. Some of this information is gathered automatically from Excel and others details are obtained through interaction with the user.

1. Spreadsheet Nature
    1. The current status of the spreadsheet; whether it is active (in use), or idle (out of use).
    2. Whether it is just for data maintaining and observing, or it incorporates input-calculations-output.
    3. The general category to which it belongs; does it hold data about people, assets, products, or any other categories.
2. Spreadsheet Importance
    1. The approximate amount of money related to it.
    2. The strategic importance; in terms of its role in reports generating for decision making purposes.
    3. Its security and safety importance.
3. Spreadsheet Complexity
    1. The spreadsheet size; number of used cells.
    2. The number of formulas.
    3. The number of names (variables).
    4. The advance tools used; such as Solver, Macros, If-statements, Nested-If, and Charts.
    5. Whether it contains comments.
4. Spreadsheet Use
    1. How old is it?
    2. How frequently is it used?
    3. When was the last time it was modified?
    4. The number of users or departments using it.
    5. Whether it is for internal use, or is transmitted externally [Chambers & Hamill, 2008].

Users are not asked to provide all of those properties, since it would be counterproductive to further burden an already busy employee. Fortunately, Microsoft Excel stores properties for each file from which the complexity questions can be answered programmatically.

To collect remaining data, users complete a short electronic survey, shown in Figure 3.

In designing this survey, it was taken into consideration that users usually do not prefer a lot of writing in documentation processes [Brace, 2013]. Therefore all the questions were designed as 'selecting from alternatives' and 'check boxes', except two pieces of information that are related to amount of money and number of users involved, where only numbers are required.

Once the user clicks the submit button, the approximate risk level for the spreadsheet will be calculated using the data entered and the automatically collected data. This data is then sent to a database called the 'Spreadsheet Properties Database' (SPD).

By generating this database, queries can be generated by different departments, in various priorities, and for several purposes, e.g. for archiving and making backups, spreadsheets can be sorted by their importance, while for auditing purposes the complexity will be taken into consideration.

Moreover, the SPD provides the flexibility to evaluate different quantification methods and decide for the best among them. For instance if the organisation thinks that the default quantification procedure does not value one aspect of their spreadsheets accordingly, the weighting of this factor can be changed to value or devalue this factor more.

The SPD can also be considered as a good indicator of the actual use of spreadsheets in the organisation's business, e.g. it is very easy to know the total amount of money related to spreadsheets, by simply summing all the data in the money field. It also can clearly indicate the nature of spreadsheets use, i.e. what level of complexity is present in the programming, how many macros in total are there? by which departments? and in which categories? etc.

### 2.1.6 Skipping

In this step, the spreadsheet nature measurements will be used for excluding two types of spreadsheets from any further consideration; the out-of-use and the data maintaining spreadsheets. While the first type is suggested to be ignored, the second can be migrated into the databases of the concerned departments. In both cases, the software tool will have no further interaction with them, except holding records of their basic metadata in the properties database.

In practice, ignoring idle spreadsheets can be achieved by assigning negative risk scores to them. In this case, they will not appear among the highest risk sheets, but their risk levels will still be readable.

### 2.2 Risk Quantification

The suggested quantification method is to calculate a Graded Point Average GPA of the three factors; Importance, Complexity, and Use.

Similar to the grading method in universities, each factor will be given a specific weight along with a relative grade, from which the GPA will be calculated using this formula:

$$GPA = \frac{\sum(grade * weight)}{\sum weights}$$

Table 1 shows these factors, their suggested weights and marking criteria. For each factor two different risk scores are calculated;

### 2.2.1   Absolute Score:

This score is calculated for each spreadsheet based on its own properties, independently from other spreadsheets within the organisation.

To do so, pre-defined thresholds for each measurement are used to assign its risk score, as shown in table 1. Obviously users will be able to calibrate and modify these thresholds via the software tool's settings in accordance with their actual needs.

The following example illustrates this quantification method:

Depending on the default definitions for the thresholds, the money measurement will be compared with two thresholds; the first one is defined as £1000, while the second threshold is £10,000. If the spreadsheet money is lower than the first threshold its risk score will be minimum (30%), if its value is between the two thresholds, the risk score will be medium (60%), and finally if its money amount is greater than the second threshold, the risk score will be maximum (100%).

Similarly, other measurements will be translated into percentage scores using specific thresholds. Each spreadsheet will have its own absolute risk score depending solely on its properties, regardless other spreadsheets available in the database. So this score does not change when new spreadsheets are added to the database, or when sheets are removed from it.

**2.2.2 Relative Score:**

On the other hand, the relative score is a dynamic figure that indicates the actual position of the spreadsheet measurement compared to all other spreadsheets in the database. Therefore, it is expected to get changed every time a new spreadsheet is added to the database, or even when any spreadsheet record is updated or deleted.

To find this score, some descriptive statistical functions are used instead of the user-defined thresholds. The chosen functions here are the quartiles;

The quartiles of a sorted set of values are defined as: the three values that divide the whole set into four equal groups [Journet, 1999].

**Table 1 Risk Quantification**

| Factor | Weight | Marking Criteria | | Grade** |
|---|---|---|---|---|
| | | Answer | | |
| | | Absolute | Relative | |
| **Importance:** | **15** | | | |
| Safety or Security Importance | 5 | Yes | | 100% |
| | | No | | 30% |
| Decision-Making Importance | 5 | Yes | | 100% |
| | | No | | 30% |
| Money Importance* | 5 | ≥ 10,000 | ≥Q3 | 100% |
| | | in between | in between | 60% |
| | | < 1,000 | <Q2 | 30% |
| **Complexity** | **15** | | | |
| Size (# of cells) | 2 | ≥ 1,000 | ≥ Q3 | 100% |
| | | in between | in between | 60% |
| | | < 100 | < Q2 | 30% |
| # of Formulas* | 3 | ≥ 100 | ≥ Q3 | 100% |
| | | in between | in between | 60% |
| | | < 10 | < Q2 | 30% |
| # of Variables* | 1 | ≥ 10 | ≥ Q3 | 100% |
| | | in between | in between | 60% |
| | | < 5 | < Q2 | 30% |
| Macros | 3 | Yes | | 100% |
| | | No | | 0% |
| Solver | 2 | same as Macros | | |
| IF statements* | 1 | same as Variables | | |
| Nested-IF | 2 | same as Macros | | |
| Charts* | 1 | ≥ 5 | ≥ Q3 | 100% |
| | | in between | in between | 60% |
| | | < 2 | < Q2 | 30% |
| Comments | | if comments are used, subtract 5% from the final complexity grade | | Optional |
| **Use** | **8** | | | |
| Frequency | 3 | Always | | 100% |
| | | Sometimes | | 60% |
| | | Rarely | | 30% |
| # of Users | 3 | same as Charts | | |
| Internal or External | 2 | External | | 100% |
| | | Internal | | 30% |
| | **38** | | | |

\* Whenever a measurement value is zero, its assigned score will also be zero.
\*\* Idle (non-active) spreadsheets will have negative risk scores.

Up to here, each spreadsheet will have a specific record in the properties database, as well as a specific percentage risk level, from which three categories can be recognised as follows:

- Risk levels of 75% and above will considered as very high risk.

- Between 50% and 75% as high risk.

- Between 25% and 50% as medium risk.

- And finally, the remaining sheets, less than 25%, will be considered as low risk.

Table 2 below shows suggested risk mitigation and reduction plans for each category:

*Table 2 Risk Reduction Policy*

| **Risk Grade** | **Description** | **Action** |
|---|---|---|
| 75% - 100% | Very High Risk | - Monitoring<br>- Auditing& Error Checking<br>- Archiving & Backup |
| 50% - 75% | High Risk | - Auditing<br>- Archiving |
| 25% - 50% | Medium Risk | - Archiving |
| less than 50% | Low Risk | - Nothing |

In the further applications, it is suggested that once a new high risk spreadsheet is created and saved anywhere in the network, an alert will be automatically sent by the software tool to the management and IT departments, so that further auditing and monitoring processes can be taken.

### 2.3 Summary

A summary of the potential risks and their mitigation methods is shown in Table 3.

*Table 3 The Relation between Risk Mitigation Processes and Potential Risks*

| Process | Potential Risks | | | | |
|---|---|---|---|---|---|
| | **Loss** | **Errors** | **Ambiguity** | **Redundancy** | **Fraud** |
| Migration& Skipping | | | | ✓ | |
| Training Sessions | | ✓ | ✓ | ✓ | |
| Documentation | | | ✓ | | |
| Auditing | | ✓ | | | ✓ |
| Monitoring | | ✓ | | | ✓ |
| Versions Archiving | ✓ | | | ✓ | |
| Backups | ✓ | | | | |
| Encryption | | | | | ✓ |

It can be clearly noticed that developing this software tool can support most of the spreadsheet risk mitigation methods by identifying high risk sheets. Figure 1 shows a flowchart of the suggested spreadsheet risk management policy and the use of the software tool within it.

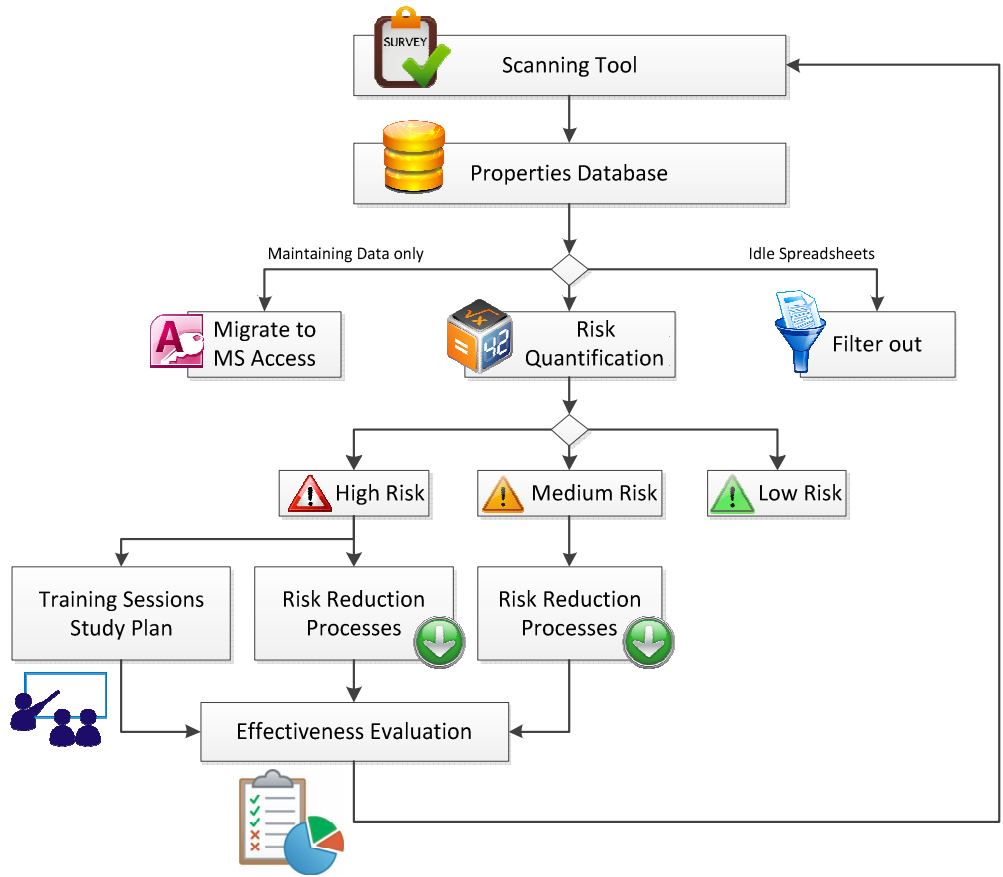

*Figure 1: Risk Management Policy Flowchart*

### 2.4 RiskCalculator Interface

Figure 2 shows the developed add-in after being installed and run in Excel.

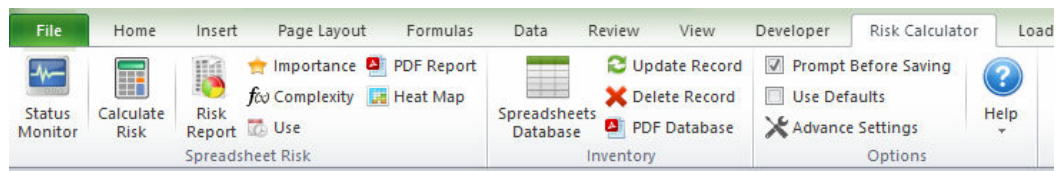

*Figure 2: Risk Calculator Ribbon as Add-in for Excel*

The 'Risk Calculator' Ribbon consists of three main groups of functions;

### 2.4.1 'Spreadsheet Risk' functions

This group gathers the functions that are solely related to each Excel workbook or file, which are: spreadsheet status regarding to the database, performing risk calculation, collecting properties for the active workbook, as well as automatically generating risk reports for the three main measurements; importance, complexity, and use. Moreover, an additional feature of representing the complexity properties graphically on the active workbook is also available via the 'heat map' command button.

**Status Monitor**

The first interaction between users and the add-in is the status monitor, with which they will be able to get a quick indication of the current situation of their spreadsheet in terms of its existence in the database, and to which extent its correlated record is accurate and up-to-date.

**Calculate Risk**

It collects the spreadsheet properties, calculates its risk levels accordingly, and stores the final outcomes along with the collected properties into a new record in the properties database. Once these processes are successfully done, the corresponding risk report will be generated and shown on the screen, and the spreadsheet status will be automatically updated. If the 'defaults' option (from the settings group) was not already chosen, the tool will ask the user to fill a short electronic survey (figure 3) once he clicks 'calculate risk'.

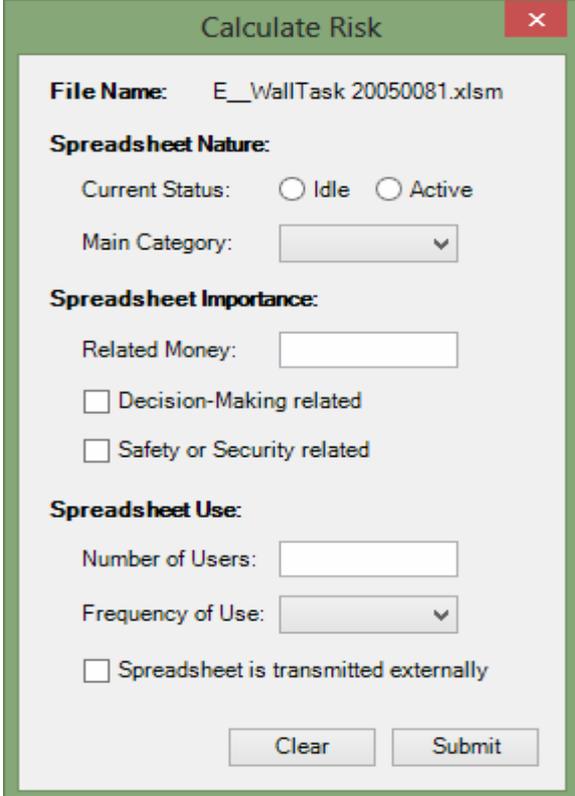

*Figure 3: RiskCalculator Survey*

### 2.4.2 Risk Report

Besides being automatically shown after a successful calculation process is executed, users can reopen the risk report anytime by clicking the risk report button. Figure 4 shows an example of this report.

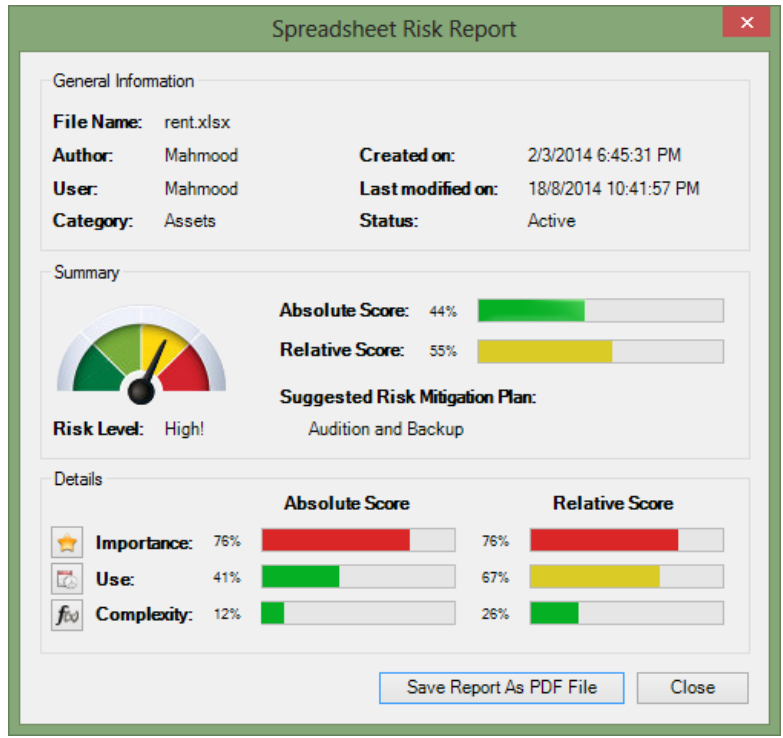

*Figure 4: Spreadsheet on-screen Risk Report*

The risk report contains general information about the spreadsheet such as file name, author, the date when it was created, last user, the nearest date it was modified on, its general category, and actual status.

In addition it includes general indications for the overall risk levels; both the absolute and relative risk scores, as well as suggested risk reduction plans that basically depend on the risk level.

The third part of this report is the detailed risk indicators for importance, use, and complexity. Users can also access a detailed report for each factor by clicking on their direct links.

Finally, besides reading the report from screen, users can also save it as a PDF file, in order to keep a soft copy of it, send it online as an email attachment, or print it out as a hard copy.

### 2.4.3 Heat Map

The 'Heat Map' function provides users with a graphical representation for the complexity measurements; by colouring cells background with various colours depending on the formulas they contain. Once the 'heat map' function is activated, a key for these colours and their corresponding meanings appears on the custom task pane in the left side of Excel's window.
Moreover, the type of the 'heat map' button has not been chosen as a normal command button but rather as a 'toggle button', so when user clicks once again on it, the function will be deactivated, undoing all its changes on cells format.

Figure 5 illustrates the function of this button on a testing spreadsheet.

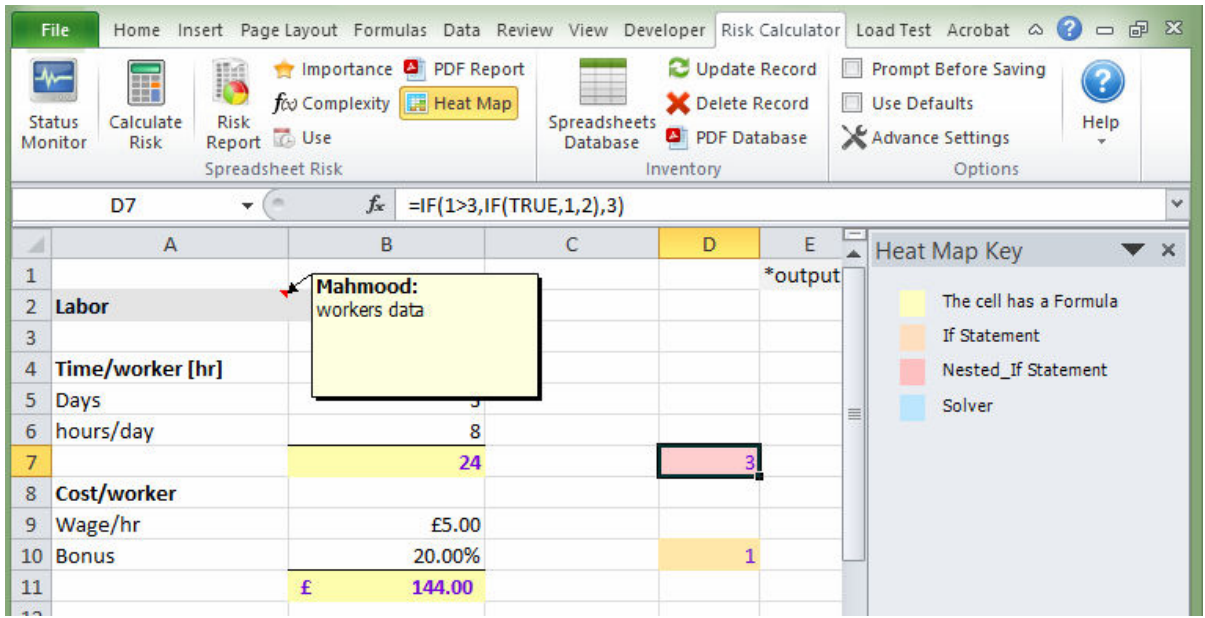

*Figure 5: Heat Map Feature*

**2.5  'Spreadsheet Properties Database' functions**

The SPD or inventory functions group is designed mainly to give users the needed control on the specific record of the active spreadsheet in the properties database, such as updating this record with the most recent data, or remove it completely from the database. Moreover, it also enables users to access the entire database with all its records, in order to view their content, or even apply any processing operations on them; such as filtering, sorting, or running any specific queries they might need. Similar to risk reports, users are also able to export the entire database into a PDF file by clicking the 'PDF Database' button.

**2.5.1  Options**

This group can be considered as the control panel for the software tool in all aspects. Users can have a full control of the add-in settings and behaviour through the options available herein. Additionally, this group gives users the needed flexibility for calibrating the risk calculator parameters. The functions provided in this group can be mainly sorted into three types;

General Behaviour Functions; which includes two quickly accessible options:

Firstly, prompting users to calculate or update their spreadsheets risk records every time they save Excel files, and secondly, using the default values when calculating risk levels. Default values are pre-defined values for all the fields that users are usually asked to fill via the properties survey. When users enable the defaults checkbox, the defaults are used to automatically fill these fields without asking users to give any information. In other words, the feature of using defaults can act as a fast alternative for the properties survey. Such methods can be used with spreadsheets of similar use and importance properties, or when these data are unknown or unimportant.

Furthermore, by using defaults, the amount of money related to the active spreadsheet will be automatically calculated either by summing up all the cells that have currency formatting, or by finding the maximum value among them. This choice can be made by users in the 'advance settings'.

Obviously, some limitations can appear here due to the reliance on currency formatted cells; as users sometimes do not use the cell's formatting properties. Testing the add-in for this particular issue, it was found that the importance risk score fell from 53% to 20% when cell formatting was changed from 'currency' to 'general'. Therefore, users are recommended to correctly use the formatting properties for each cell before using the 'defaults' option.

When the add-in is installed into Excel for the first time, both checkboxes; the prompting and defaults are unchecked, and the default option for the money field is the maximum value by default.

### 2.6.1  Advanced Settings

The advanced settings are categorized as follows:

**Defaults Values** – users are able to define the default values to be used instead of the survey when this feature is enabled.

As shown in figure 6, these data are related to spreadsheet nature, use and importance properties. It is very similar to the survey, however, for the 'money field' users can choose either the summation or the maximum value among the currency-formatted cells to be used as a default value.

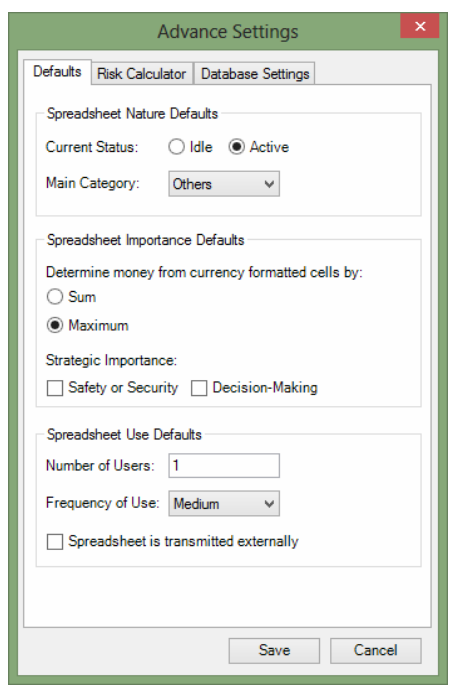

*Figure 6: Advance Settings - Defaults*

**The 'Risk Calculator Settings'**; which includes the weights of risk factors, the score levels, and the thresholds to be used for calculating absolute risk scores. Figure 7 shows a screenshot of the risk calculator settings window.

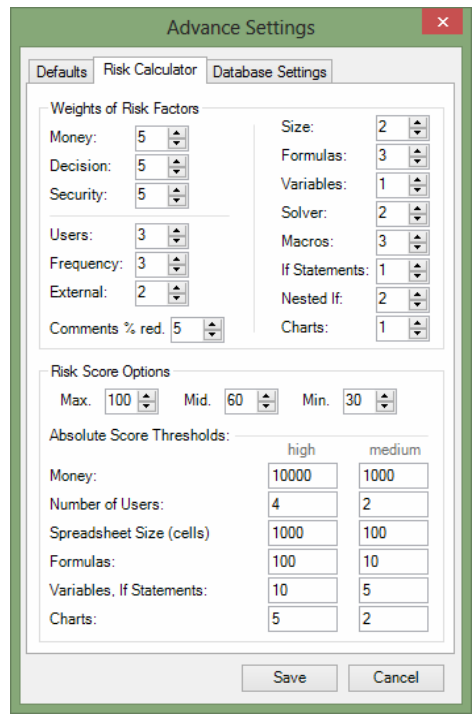

*Figure 7: Advance Settings - Risk Calculator*

Through this screen, users can calibrate the weights of risk factors depending on their actual spreadsheets' use. For example, in logistics and supply chain management departments, where operations research calculations are usually done for finding optimal production decisions, the use of the Solver add-in is considered more significant. Therefore, users from such departments can easily increase the 'Solver' weight as a complexity risk factor via this screen. In this case, spreadsheets that contain Solver operations will have more complexity scores than those which do not.

**'Database Settings'**; which contains:

- The time period threshold for a spreadsheet to be considered as 'out-of-date'.
- The Database security options, such as setting and changing its password, and controlling its opening options; making it open as a webpage instead of an Excel file.
- The option of updating the entire database and recalculating the risk scores for every single record it contains.
- The option of removing all the records of files that are deleted from the hard drive.
  Figure 8 shows all these options.

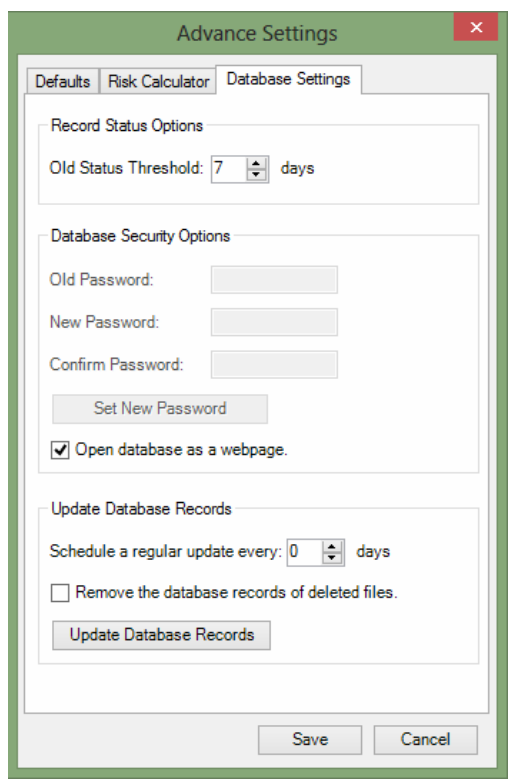

*Figure 8: Advance Settings - Database Settings*

**2.7   Testing and Analysis**

A sample of the EUSES spreadsheet corpus was used for testing processes. The results of these tests have been analysed and used for calibrating and improving the software tool functions.

Having a brief look into the corpus, it contains 690 functional spreadsheets that are related to financial operations, which have all been used for testing the 'Risk Calculator Add-in'.

Since these Excel files have not been designed by the researcher, most of the use and importance measurements could not be determined, such as the number of users, frequency of use etc. For this reason the 'default values' approach have been used to deal with these measurements. This made the testing process faster, and its results were basically built on the complexity and money measurements.

Table 4 summarizes the test outcomes:

*Table 4: Summary of Test Outcomes*

| | |
|---|---|
| **Number of Excel files** | **690** |
| **Total Time needed** | **38.74 minutes** |
| **Average Time** per file | **3.4 seconds** |
| **Average Money** | **$74,944,542.72** |
| **Average Spreadsheet Size** (number of cells) | **775.3** |
| **Oldest file created in** | 1995 |
| **Newest file created in** | 2004 |
| **Average Relative Risk Score** | 34.9% |
| **Min. Relative Risk Score** | 26.8% |
| **Max. Relative Risk Score** | 69.5% |
| **Average Absolute Risk Score** | 30.5% |
| **Min. Absolute Risk Score** | 21.3% |
| **Max. Absolute Risk Score** | 62.1% |

This testing operation had many benefits, it helped in detecting and fixing some bugs in the add-in, it also supported the calibration of the risk factors' weights, and it opened the door for further capabilities and analysis possibilities.

Moreover, some compatibility issues have been discovered and solved. It gave us some indications to the differences in the software architecture and internal definitions used in the old and new versions of Microsoft Excel.

### 2.8 Further Analysis Capabilities

One of the most interesting outcomes of this research is the wide possibilities that the 'risk calculator' add-in along with the 'properties database' can provide.

With its valuable numerical data sets, the 'properties database' can open the door to further study and analysis of the spreadsheets nature as well as the human use of it. Such analyses can lead to interesting and sometimes unexpected outcomes. It also reveals some hidden relations between factors and aspects that we have never studied before, or were not able to observe.

This subsection introduces examples for such data analysis for the 690 financial Excel sheets that are tested in this stage.

#### 2.8.1 The relation between spreadsheet size and number of formulas:

As shown in the following chart (figure 9) these two factors are directly correlated; as the largest the spreadsheet is, the highest possibility to have formulas it has.

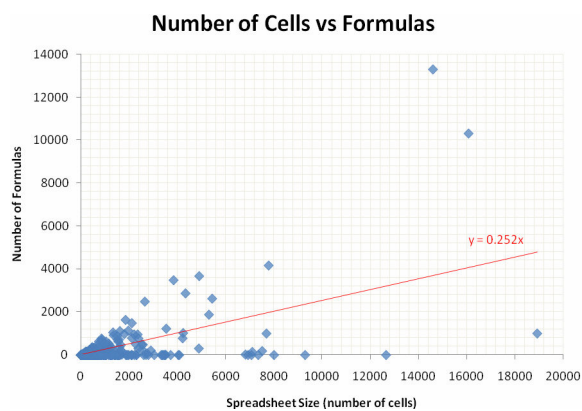

*Figure 9: Cells vs. Formulas Graph*

Since this direct correlation between spreadsheet number of cells and number of formulas can be considered as self-evident, it has never been studied in more detail before. However, understanding such relations might lead to a deeper understanding of the actual interaction between users and spreadsheet applications, and thus, it might help in improving the efficiency of these applications and reducing their risks.

Based on the results shown in figure 10, a further step towards understanding such relation as well as studying the human factors of using and designing spreadsheets can be taken, as shown in the graph, formulas occupies 25% of spreadsheet's cells.

**2.8.2 The relation between the spreadsheet size and the tools used within it:**

Interestingly, the data analysis of this test operation shows the following outcomes for the relation between Excel sheet size and the use of If statements and Charts in it.

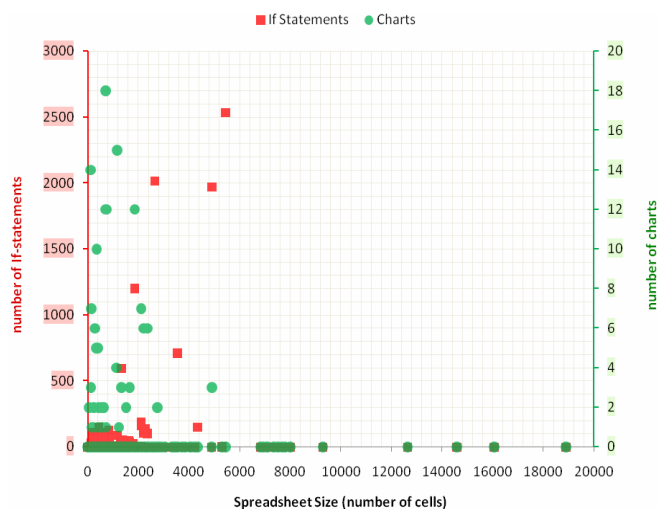

*Figure 10: Cells vs. Charts and If-statements Graph*

As shown in figure 10, 'If-statements' are mostly used within the mid-size spreadsheets, that have between 1,500 and 5,500 cells. On the other hand, charts can be found in the small-size spreadsheets (of 10 to 2,500 cells size). It was also observed that both tools are not usually used in the large-size spreadsheets (of more than 5,500 cells size). Considered as unaddressed research perspective before, such analysis can give a deeper understanding of the actual human use of spreadsheets, which can support the scientific research on human interaction with end-user computing applications, spreadsheets errors' classification, and complexity taxonomies. It can also draw the attention to the powerfulness of the add-in developed here in collecting and building multidimensional models of several spreadsheet complexity factors, studying their relative trends, keeping records of them, as well as giving users the desired flexibility to tune up their weights in order to form an overall risk score.

**2.8.3 The relation between spreadsheets size and amounts of money they contain:**

The following graph (figure 11) shows that the large money amounts are usually included in the small-sized spreadsheets (of cells less than 2,500), while no money amounts can be found in the large spreadsheets. Based on these observations, it can clearly be seen the effectiveness of having different risk measurements for both spreadsheet's complexity and its financial importance. As large money amounts are related to small spreadsheets, the importance score of such sheets will be relatively high, while their complexity scores might be low. Having two different measurements, both risk indicators can be easily detected, and thus properly dealt with.

Point X in the graph below is a good example. As the size of this spreadsheet is only 360 cells, with no macros, no solver or any if statements involved, its relative complexity score is 33.7% (low compared to the max. 86.7%). On the other hand, containing large money amounts of about $469 Millions, its relative importance score was at the maximum level of 53.3% (this is the maximum importance level because no decision-making or safety importance is considered by default). Consequently, this spreadsheet had a high overall risk score of 50.5%. The powerfulness of the developed add-in can be obviously seen here, since the high risk incorporated with this spreadsheet was detected, despite its low complexity.

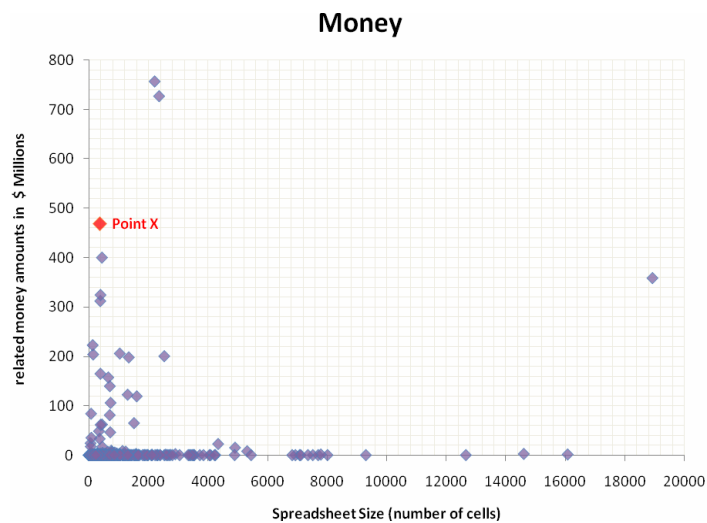

*Figure 11: Cells vs. Money Graph*

## 3    CONCLUSIONS

End-User Computing Applications can be a nightmare for any organisation, as their uncontrolled use encounters many potential risks. However, a good understanding of these risks, and applying suitable policies for managing them, can convert spreadsheets into reliable tools that aid the organisation in reaching its goals, benefiting from all the advantages they have.

This research has introduced and developed a software tool that automates spreadsheets risk management, by recording their properties in a special database, and assigning risk scores to them based on their importance, use, and complexity. Consequently, further auditing processes can be applied on the high risk sheets only. Such a method saves much time, effort, and money, and can be considered as bridging the gap in the other risk mitigation methods.

## 4    FUTURE WORK

Further improvements and research fields can be built on this research, such as:

- Centralizing the developed tool into having a single database within the entire organisation, using a Microsoft SharePoint server.

- Developing a mechanism for generating and sending alerts to management once any high risk spreadsheet is saved or modified within organisation's servers.

- Improving the complexity factors by differentiating between Excel's built-in functions, and user-made formulas.

- Using the developed add-in for deeper studying of the relations between spreadsheet internal properties and used tools, as well as the human attitudes in designing and using spreadsheets.

**REFERENCES & BIBLIOGRAPHY**